# Impact of OH-groups on the mobility of linkers and guests in UiO-66 (Zr)

Alexander E. Khudozhitkov,[ab] Sergei S. Arzumanov,[a] Daniil I. Kolokolov,*[a] and Alexander G. Stepanov*[a]

[a]*Boreskov Institute of Catalysis, Siberian Branch of Russian Academy of Sciences, Prospekt Akademika Lavrentieva 5, Novosibirsk 630090, Russia.*
[b]*Novosibirsk State University, Pirogova Street 2, Novosibirsk 630090, Russia*
e-mail: kdi@catalysis.ru; stepanov@catalysis.ru

**Abstract**

UiO-66 (Zr) is a metal-organic framework known for its thermal and chemical stability and wide range of possible applications. In particular, this material has high separation selectivity of various hydrocarbon mixtures. Moreover, it is was shown that the performance depends on the hydroxylation state of material (reversible phase transition $Zr_6O_4(OH)_4 \to Zr_6O_6$ at 250 – 300 °C). Despite all the attention the UiO-66 received over past few years, the impact of the dehydroxylation on its properties remains poorly understood. In this contribution we apply $^2$H NMR experimental method in order to compare the mobility of butane isomers in hydroxylated and dehydroxylated form of UiO-66. We provide estimations of the translational diffusion coefficients and show which hydroxylation state of the material has higher separation selectivity. Moreover, the impact of the hydroxylation on the structural mobility of UiO-66 is discussed. Correlation between parameters of structural dynamics and the diffusivity of guest molecules indicates that for the microporous material with small windows even subtle changes can lead to a tangible improvement of the properties.

## 1. Introduction

UiO-66 (Zr) is a metal-organic framework (MOF) comprised of zirconium oxide clusters ($Zr_6O_4(OH)_4$) connected with 1,4-benzodicarboxylic acid linkers that build up a 3D porous network formed by tetrahedral (effective diameter $d = 8$ Å) and octahedral ($d = 11$ Å) cages interconnected by narrow windows ($d \sim 6$ Å). This material is well-known for its high thermal stability, catalytic properties[1-3] and possibility to exploit this material in adsorption and separation applications[4-6]. UiO-66 (Zr) is one of the most studied MOFs (ca. 3000 papers). However, there are some interesting questions that have not been regarded so far.

Upon heating the material to 250 – 300 °C the metal oxide undergoes reversible structural changes $Zr_6O_4(OH)_4 \rightarrow Zr_6O_6$ and the dihydroxylation of UiO-66 occurs [7]. However, the impact of the dehydroxylation on the properties of UiO-66 (Zr) remains poorly studied. Bambalaza et al. show that hydroxylated UiO-66 has higher uptake capacity of hydrogen than dehydroxylated form [8]. Similar effect was demonstrated for $CH_4$ and $CO_2$ adsorption. Using neutron scattering the authors prove that oxygen –O and hydroxyl group –OH are the main adsorption sites for methane and carbon dioxide. With a density functional theory (DFT-D) calculation they prove that the binding energy to the hydroxyl group is slightly higher [9]. Using infra-red (IR) spectroscopy Wiersum et al. confirm that there is only physisorbed $CO_2$ species regardless of the hydroxylation and adsorption enthalpy in hydroxylated form is 6 – 10 kJ mol$^{-1}$ higher [10].

Grissom et al. applied IR spectroscopy in order to compare the diffusivity of benzene and alkylaromatics in hydroxylated and dehydroxylated form of UiO-66 (Zr) [11]. The experiments revealed that diffusion of *p*-xylene occurs slower in the hydroxylated form ($D_{hydro}/D_{dehydro} \approx 70$). Interestingly, the activation barrier of diffusion for *p*-xylene in the hydroxylated form 42.7 kJ mol$^{-1}$ remains within the experimental uncertainty with the diffusion barrier in the dehydroxylated form 46.9 kJ mol$^{-1}$. This result implies that the breaking of the hydrogen bond between OH-group and the adsorbed molecules is not the limiting step of diffusion. Instead, the activation barrier arises during the window crossing by the guest molecules.

So, the hydroxylation form of UiO-66 is known to have impact on the adsorption properties and the diffusivity of the hydrocarbons. The main reasons of this impact are the structure deterioration (if the dehydroxylation is performed at high temperature) and the possibility of forming hydrogen bonds in the hydroxylated form of UiO-66. However, there is the other possible outcome of the dehydroxylation that was not discussed previously.

The organic linkers of UiO-66 were shown to have high rotational freedom that leads to the fluctuation of the window aperture between adjacent cages [12-13]. The structural mobility increases the range of the molecules that can be adsorbed in UiO-66. As a result, such molecules as xylenes can be introduced inside the material despite having kinetic diameter exceeding the nominal window size (6 Å). Therefore, if the hydroxylation affects the rotational motion of the organic linkers it may impact as well the diffusivity of adsorbed species.

In this contribution we perform detailed characterization of organic linker mobility in the guest-free hydroxylated form of UiO-66 employing $^2$H NMR spectroscopy. The dynamics of butane isomers in the hydroxylated form of UiO-66 is studied as well. We assess the activation barrier of diffusion and diffusivity. The results are compared with the data from our previous works where the same experiments were performed in the dehydroxylated form of UiO-66. Comparison of linker mobility reveals the difference between "gate-opening" effect in both hydroxylation forms. This difference impacts the butane diffusion and makes the performance of UiO-66 dependent on the structural form.

## 2. Experimental section

### 2.1. Materials

The synthesis and activation of UiO-66 (Zr) material were performed according to the previously reported procedure[14]. The material was synthesized using deuterated terephthalic acid as well in order to study the mobility of linkers. The hydrogenated form of UiO-66 (Zr) was used to study the mobility of guest molecules. The resulting material represents crystalline agglomerates with the average crystallite size of 500-1000 nm (Figure 1), good crystallinity (Figure 2), and a Brunauer–Emmett–Teller specific surface area $S_{BET}$ = 1145 m$^2$/g (via N$_2$ sorption on a dehydroxylated form, activated at 250 ºC, Figure 3).

Butane isomers, *n*-butane-$d_{10}$ and isobutane-$d_1$, were purchased from Sigma-Aldrich, Inc. and used further without additional purification.

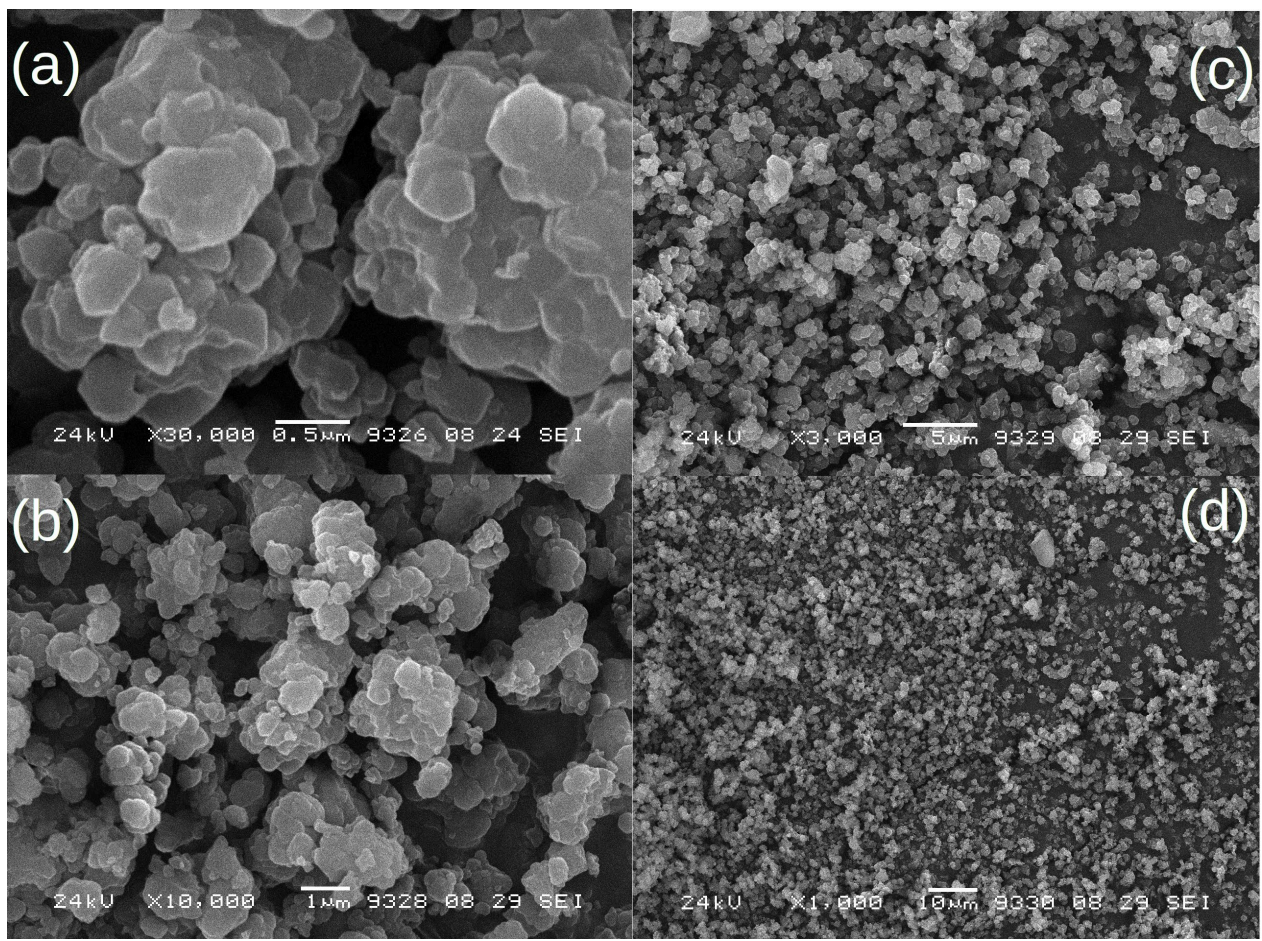

**Figure 1**. (a-d) SEM images of the synthesized and dried UiO-66 compound.

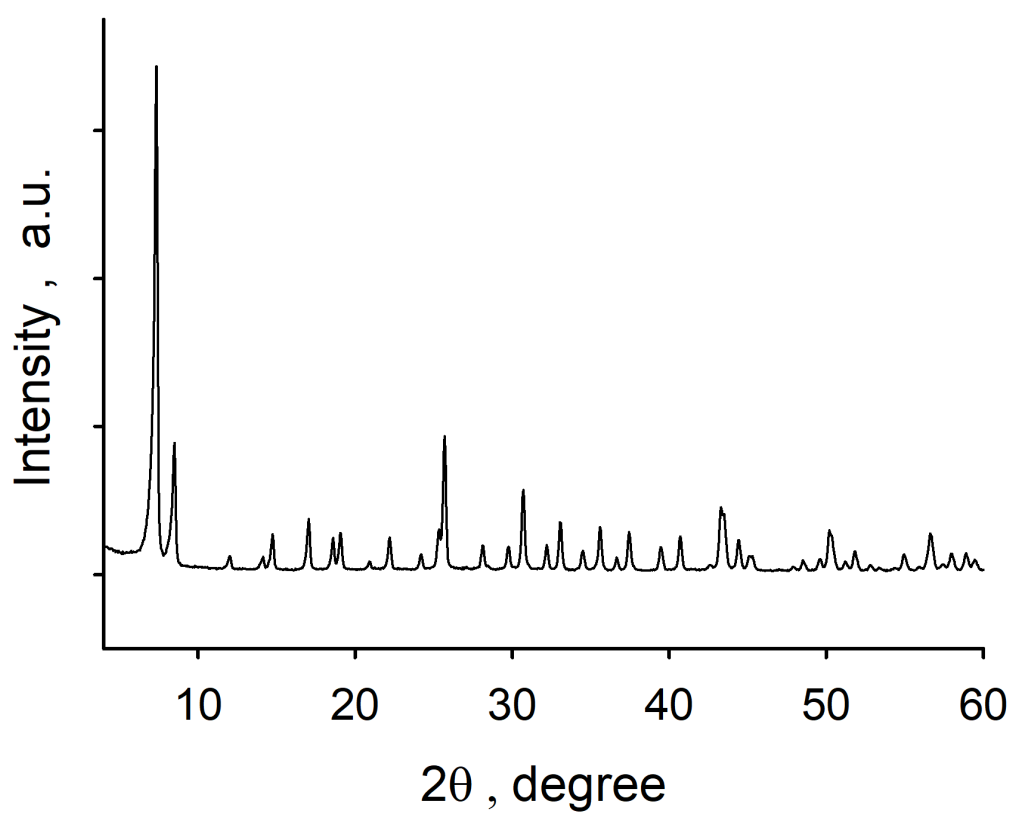

**Figure 2**. XRD patterns of the synthesized and dried UiO-66 compound.

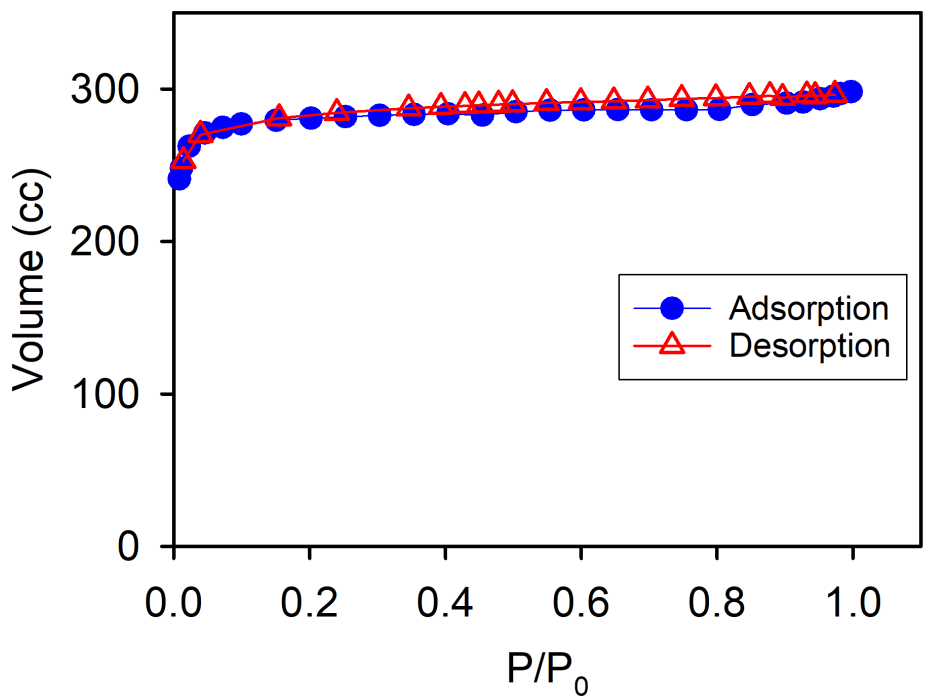

**Figure 3**. Nitrogen adsorption isotherms and textural characteristics for for UiO-66 (Zr) activated at 250 °C. $S_{BET}$ = 1145 m$^2$/g, $S_{ext}$ = 33 m$^2$/g, $V_{total}$ = 0.46 cm$^3$/g, $V_{micro}$ = 0.42 cm$^3$/g.

**2.2. Sample preparation**

Preparation of the samples for NMR experiments was performed in the following manner. The powder of UiO-66 (Zr) (~0.07 g) was placed into a special glass cell of 5 mm diameter and 3 cm length. Then, the cell was connected to the vacuum line and the sample was activated at 363 K for 24 h under vacuum in order to obtain the hydroxylated form of material. After cooling the sample back to room temperature, the material was exposed to the deuterated alkane gas (5 mBar, in case of *n*-butane and 36 mBar for isobutane) in the calibrated volume of 52 cm$^3$. The residual quantity of the guest not consumed by the material (1– 5 mBars) was finally frozen with liquid nitrogen on the sample. The quantity of the adsorbed guest was regulated by the gas pressure inside the calibrated volume. The adsorption quantity is 1 molecule/u.c. for *n*-butane-$d_{10}$ and 9 molecule/u.c. for isobutane-$d_1$ (1 u.c. = 4 octahedral cages + 8 tetrahedral cages). The amount of adsorbed isobutane-$d_1$ was chosen larger because of the lower quantity of deuterium in the molecule than in the molecule of n-butane-$d_{10}$. However, at such loading, the number of cages containing more than one adsorbate is still negligible. After adsorption, the neck of the tube was sealed off with a micro-torch flame, while the material sample was held in liquid nitrogen to prevent its heating by the flame. Prior to NMR investigations, all sealed samples were kept at 363

K for 24 h to allow an even redistribution of the guest molecules over the porous material. Guest-free sample of deuterated UiO-66 (Zr) was activated in the same manner.

### 2.3. $^2$H NMR experiments

$^2$H NMR experiments were performed at the Larmor frequency $\omega_0/2\pi = 61.424$ MHz on a Bruker Avance-400 spectrometer using a high-power probe with a 5 mm horizontal solenoid coil. All $^2$H NMR spectra were obtained by a Fourier transform of a quadrature-detected and phase-cycled quadrupole echo after two phase-alternating 90°-pulses in the pulse sequence (90°$_x$ − τ − 90°$_y$ − τ − acquisition− t), where τ = 20 μs and t is a repetition time of the sequence during accumulation of the NMR signal. The duration of π/2 pulse was 1.8 μs. Spectra of the guest molecules were typically obtained with 4−100 scans with a repetition time ranging from 0.5 to 15 s. Spectra of the deuterated guest-free UiO-66 (Zr) were obtained with 30000-60000 scans with a repetition time 1-2 s. Inversion−recovery experiments for measurements of spin−lattice relaxation times ($T_1$) were carried out using the pulse sequence 180°$_x$ − $t_v$ − 90°$_x$ − τ − 90°$_y$ − τ − acquisition − $t$, where $t_v$ was the variable delay between the 180° and the 90°-pulses. Spin-spin relaxation time ($T_2$) was measured by a Carr-Purcell-Meiboom-Gill pulse sequence. The repetition time $t$ was always longer than 5-fold of the estimated relaxation time $T_1$.

The temperature of the samples was controlled with a flow of nitrogen gas by a variable-temperature unit BVT-3000 with a precision of about 1 K. The sample was allowed to equilibrate at least 15 min at the temperature of the experiment before the start of the NMR signal acquisition. Modeling of spin relaxation time was performed with a homemade FORTRAN program based on the standard formalism [15].

### 3. Results and Discussion

### 3.1. The mobility of linkers

The influence of the OH-group on the mobility of UiO-66 linkers was studied by the analysis of the $^2$H NMR spectra line shape. Figure 4a shows spectra of the deuterated guest-free UiO-66 (Zr) in a hydroxylated form. Spectrum evolution with a temperature increase is typical for the model of π-flips around the symmetry axis of terephthalic linker (Fig.4c). Additional narrowing of the spectrum upon the temperature increase occurs and can be associated with fast ($k_{lib} \gg Q \approx 10^5$ Hz) low-angle librations of the linker around the same axis. The motional model is in line with our previous study of the linker mobility in the dehydroxylated form of UiO-66.[12] However, the simple model of π-flips with a single rate constant cannot reproduce experimental spectra even after including the impact of librations. The high number of defects typical for the UiO-66 leads to the inhomogeneity of the linker surrounding and leads to the distribution of the flipping rate $k_{flip}$. The signals of linkers with various flipping rate are superimposed resulting in a complex line shape that has features typical for slow ($k_{flip} \sim 10^3 - 10^4$ Hz) and fast ($k_{flip} \sim 10^5 - 10^6$ Hz) motion. The log-normal distribution of the flipping rate allows reproducing the experimental spectra.

$$\rho(k_{flip}) = \exp^{-(lnk_{flip}-lnk)^2/2\sigma^2}$$

The width of distribution is characterized by the parameter σ. Relatively good fitting of the experimental data can be performed with σ = 3 – 5. The center of the rate constant distribution $k$ shifts to the higher rates at elevated temperature. The temperature dependence of the distribution center position follows the Arrhenius law as shown in Figure 5. Increase of the σ value is accompanied by the increase of the preexponential factor and insignificant decrease of the activation energy (Figure 6, Table 1). Therefore, whereas the comparison of the activation energy is straightforward, the comparison of the rate constant value should be performed at the same σ value. So, the σ = 5 was chosen in order to compare results for hydroxylated and dehydroxylated form of UiO-66. The temperature dependence of the flipping rate constant for both forms is shown in Figure 5a. The experimental spectra (taken from our previous work) [12] for dehydroxylated form were remodeled with a fixed value σ = 5 that yielded the same activation energy and higher pre-exponential factor.

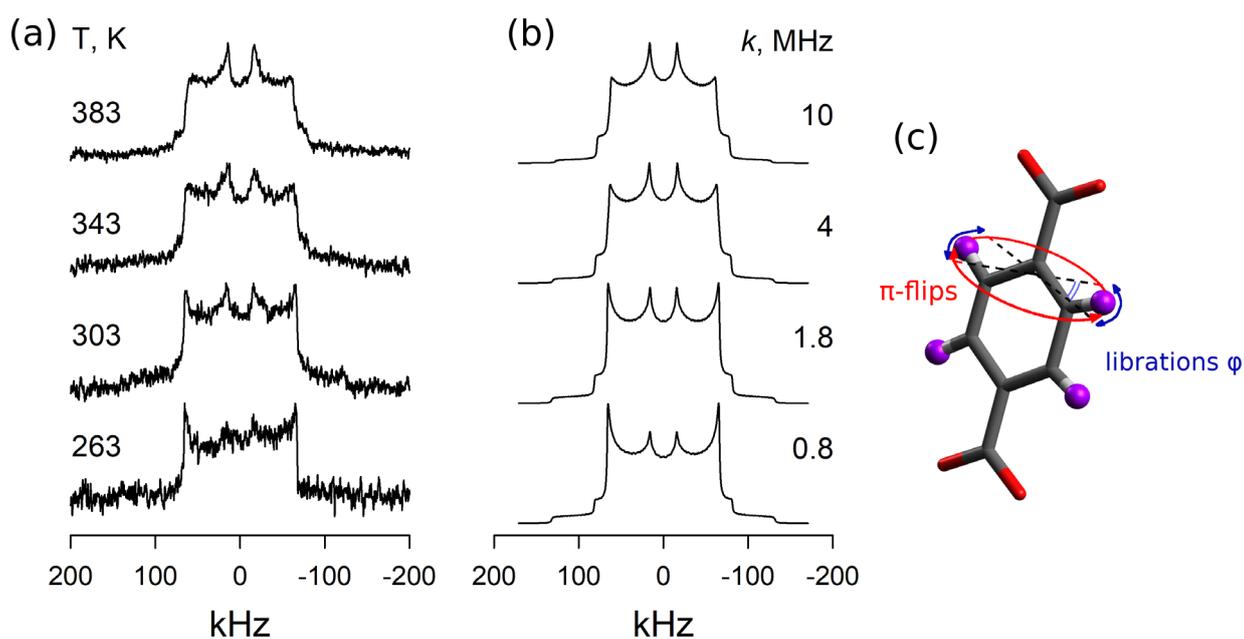

**Figure 4.** (a) Experimental and (b) simulated $^2$H NMR spectra of deuterated guest-free UiO-66 in a hydroxylated form at 263-383 K temperature region (σ = 5). (c) Schematic representation of linker motional model.

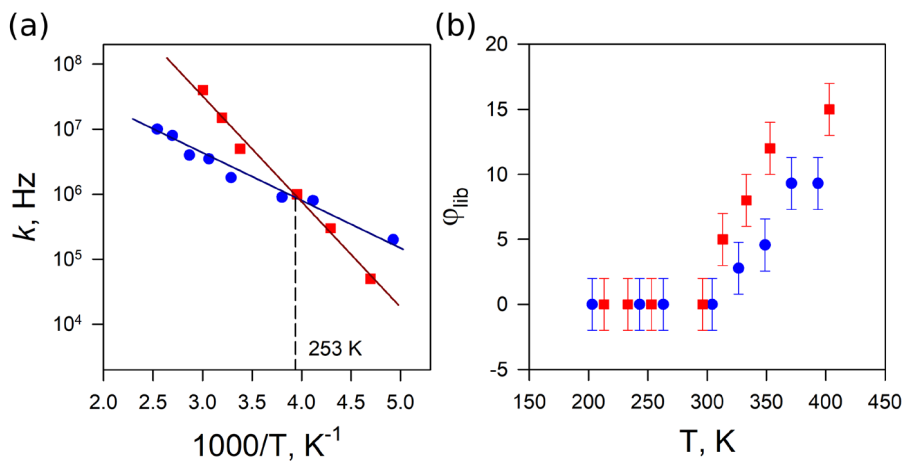

**Figure 5.** (a) Arrhenius' plot of linker flipping rate constant in hydroxylated (•) and dehydroxylated (■) form of UiO-66 (Zr). Solid lines show the fitting. (b) Temperature dependence of libration angle in hydroxylated (•) and dehydroxylated (■) form of UiO-66 (Zr).

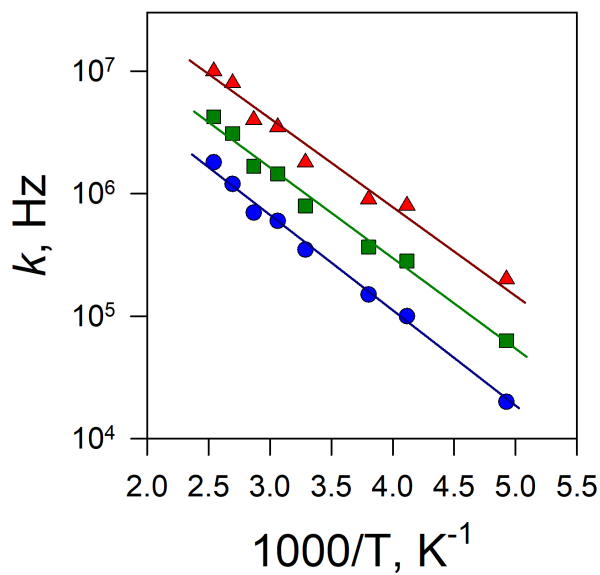

**Figure 6.** Arrhenius' plot of the flipping rate constant derived from the modelling with σ=3 (•), σ=4 (■) and σ=5 (▲).

**Table 1.** Activation energy and pre-exponential parameter of flipping motion derived at various σ.

|  | $E_a$, kJ mol$^{-1}$ | $k_0$, Hz |
| --- | --- | --- |
| σ = 3 | 15.1 ± 0.5 | $1.5 \times 10^8$ |
| σ = 4 | 14.1 ± 0.7 | $2.7 \times 10^8$ |
| σ = 5 | 13.2 ± 0.9 | $4.8 \times 10^8$ |

The mobility of organic linkers in hydroxylated form is characterized by a significantly lower activation barrier ($E_a^{hydro}$ = 13.2 kJ mol$^{-1}$, $E_a^{dehydro}$ = 30.3 kJ mol$^{-1}$) and pre-exponential factor ($k_0^{\text{hydro}} = 4.8 \times 10^8$ Hz, $k_0^{\text{dehydro}} = 1.6 \times 10^{12}$ Hz). Such relation between dynamics parameters leads to a faster flipping motion in the dehydroxylated form above 253 K. The librational motion remains in the fast-motional limit for both forms of UiO-66. However, the amplitude of librational motion in the hydroxylated form of UiO-66 is lower, that should lead to the slightly smaller window aperture in this form.

Basing on the analysis of the terephthalate linker mobility we can conclude that lower activation barrier of flipping motion in the hydroxylated form should facilitate the diffusion of the adsorbed species. At the same time, lower amplitude of libration, lower flipping rate above 253 K in the hydroxylated form of UiO-66 and the possibility of forming hydrogen bond between guest molecule and OH-group are the factors that encourage faster diffusion in the dehydroxylated form.

### 3.2. The mobility of butane isomers in UiO-66

Having compared the dynamics of organic linkers in two forms of guest-free UiO-66, we decided to look how our findings can be applied to the butane separation. The mobility of *n*-butane and isobutane in the dehydroxylated form was reported earlier [16]. In this contribution we repeated the experiment for the hydroxylated form of UiO-66 (Zr).

The spectrum of *n*-butane-$d_{10}$ consists of Lorentzian signal with a full width at half maximum (FWHM) 1 – 4 kHz and two anisotropic Pake-powder-like doublets with $Q_{\text{eff1}}$ = 15 – 24 kHz, $\eta_1$ = 0.18 – 0.5 and $Q_{\text{eff2}}$ = 36 – 46 kHz, $\eta_2$ = 0.1 – 0.15 at 113-173 K that correspond to $CD_3$ and $CD_2$ groups, respectively (Figure 7,8). The spectrum of isobutane-$d_1$ is represented by a Lorentzian signal with FWHM 7 – 14 kHz and broader anisotropic Pake-powder-like doublet signal with effective quadrupole coupling constant (QCC) $Q_{\text{eff}}$ = 43 – 67 kHz and asymmetry parameter $\eta_{\text{eff}}$ = 0.3 – 0.47. Coexistence of two spectral patterns, with isotropic and anisotropic line shapes, can be rationalized with a presence of two dynamically different states of adsorbed species that exchange slowly (ca. 50 kHz). Some guest molecules exhibit hindered dynamics (the dynamic state I) that leads to the anisotropic line shape and others can rotate isotropically (the dynamic state II) resulting in the Lorentzian signal. In our previous work we ascribed dynamical state I to the molecules localized in the tetrahedral cages and dynamical state II to the molecules occupying octahedral cages.

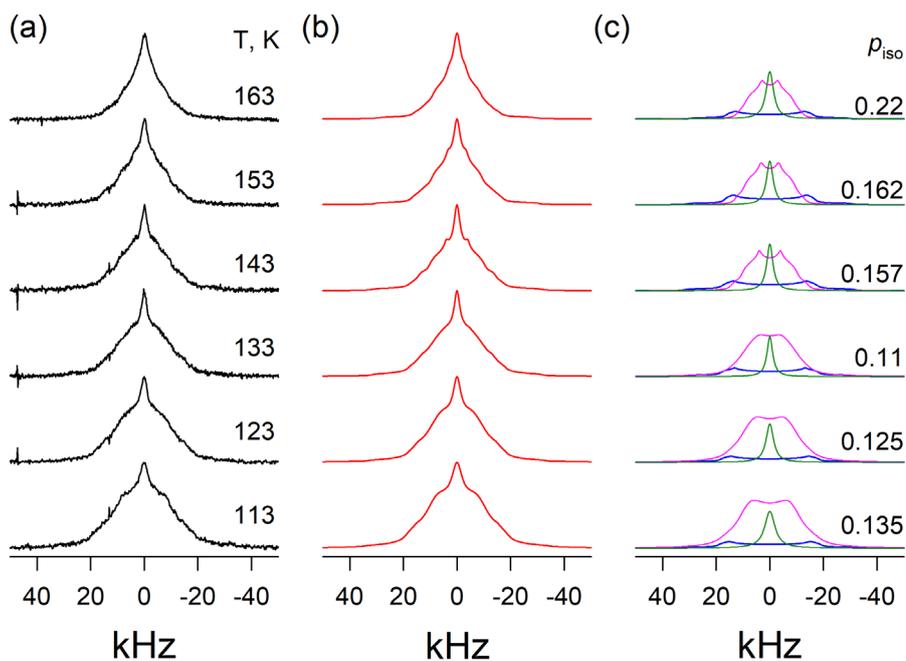

**Figure 7.** $^2$H NMR spectra of *n*-butane-$d_{10}$ in hydroxylated form of UiO-66 (Zr). (a) experimental spectra, (b) simulated spectra, (c) deconvolution of spectra to the signals of CD$_2$-group (blue) and CD$_3$-group (pink) in dynamical state I and isotropic signal (green) corresponding to the dynamical state II.

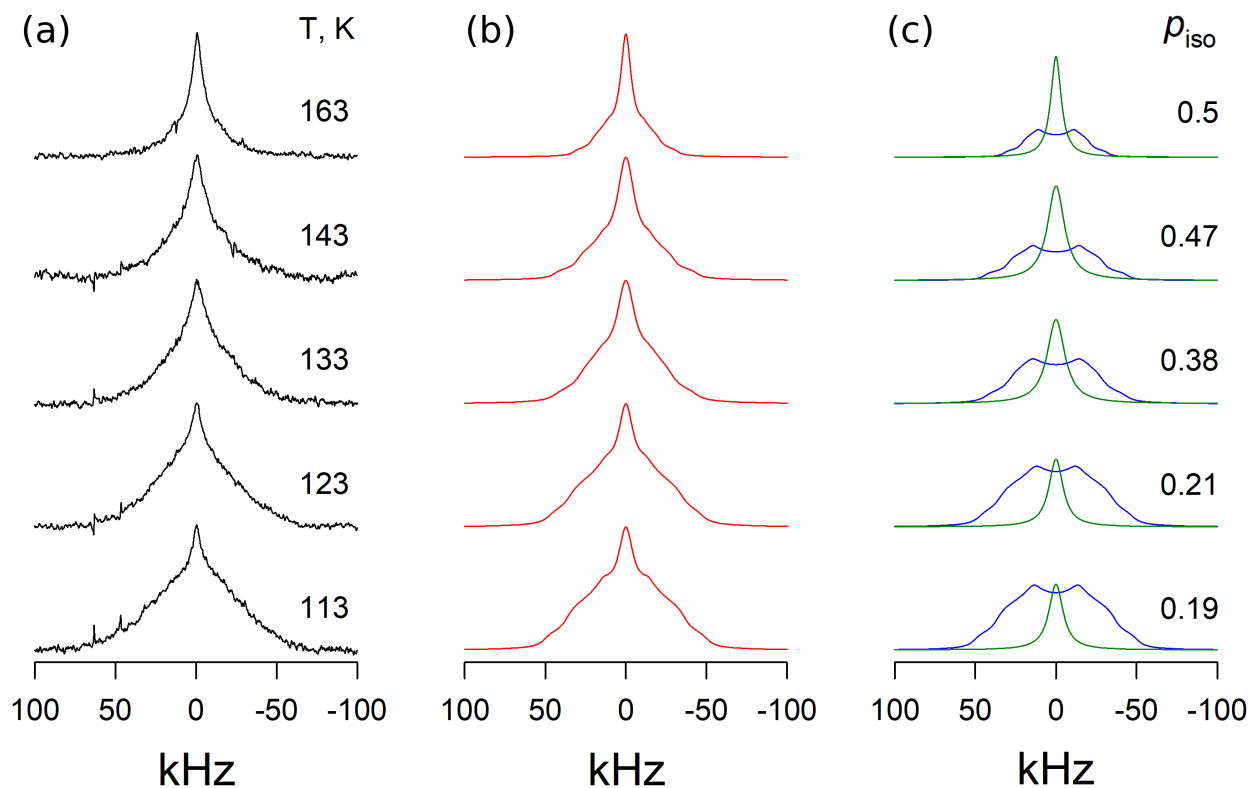

**Figure 8.** $^2$H NMR spectra of isobutane-$d_1$ in hydroxylated form of UiO-66 (Zr). (a) experimental spectra, (b) simulated spectra, (c) deconvolution of spectra to the anisotropic signal from dynamical state I (blue) and isotropic signal (green) corresponding to the dynamical state II.

The comparison of $^2$H NMR spectra collected at 113 K for hydroxylated and dehydroxylated form shows that the hindered mobility in the dynamical state I is practically the same for both forms since spectra are characterized with close effective parameters $Q_{eff}$ and $\eta_{eff}$ (Figure 9). However, the intensity of the isotropic signal (proportional to the population of dynamical state II $p_{II}$) drops down significantly upon hydroxylation of UiO-66 ($p_{II}$ drops from 0.26 to 0.13 in case of n-butane and from 0.55 to 0.19 for isobutane). Temperature increase leads to the increase of the dynamical state II population $p_{II}$. If we analyze the temperature dependence of equilibrium constant $K_{eq} = p_{II}/p_I$ (Figure 10), the enthalpy $\Delta H$ and entropy $\Delta S$ difference between dynamical states can be assessed using van't Hoff law: $ln\ K_{eq} = \Delta S/R - \Delta H/RT$. The enthalpy difference ($\Delta H_{isobut} = 5.3$ kJ mol$^{-1}$ and $\Delta H_{n\text{-}but} = 1.6$ kJ mol$^{-1}$) in the hydroxylated form of UiO-66 remains almost the same as in the case of dehydroxylated form, whereas the entropy difference ($\Delta S_{isobut} = 34$ kJ mol$^{-1}$K$^{-1}$ and $\Delta S_{n\text{-}but} = -2$ kJ mol$^{-1}$ K$^{-1}$) drops down significantly (Table 2). Therefore, the equilibrium shifts to the dynamical state I in the hydroxylated form due to the decrease of the rotational freedom in the octahedral cage instead of the energetic preference of the tetrahedral cage.

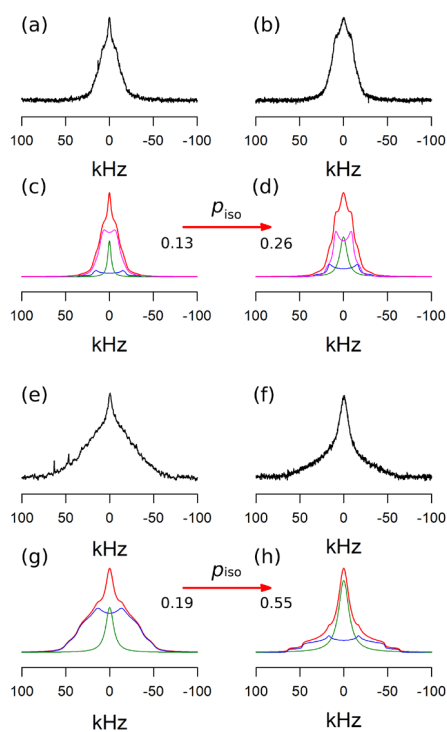

**Figure 9.** $^2$H NMR spectra of (a-d) n-butane-$d_{10}$ and (e-h) isobutane-$d_1$ at 113 K adsorbed in UiO-66 (Zr). Experimental spectra (a) and (e) correspond to the hydroxylated form of UiO-66 (Zr), experimental spectra (b) and (f) are taken for dehydroxylated form of UiO-66 (Zr). Simulated spectra (red) are presented under the corresponding experimental spectrum ((c) and (g) – hydroxylated form, (d) and (h) – dehydroxylated form). Deconvolution of experimental spectra is shown with blue (anisotropic signal for isobutane and $CD_2$-group signal for *n*-butane), pink ($CD_3$-group of *n*-butane) and green (isotropic signal) lines.

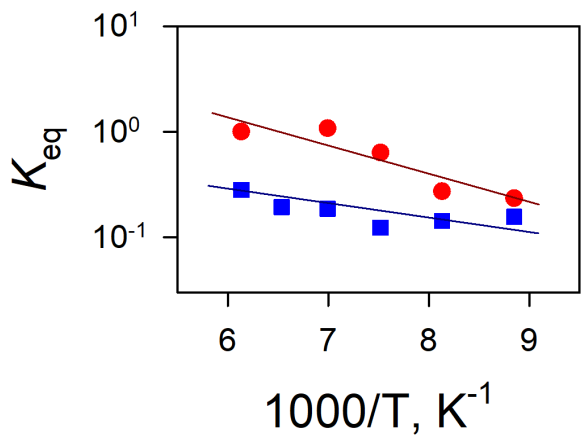

**Figure 10.** Van't Hoff plot for equilibrium constant between dynamical states I and II $K_{eq} = p_{II}/p_{I}$. Populations are derived from the spectra line shape analysis for *n*-butane (■) and isobutane (●).

**Table 2.** Enthalpy $\Delta H$ and entropy $\Delta S$ difference between dynamical state II and I derived from the $^2$H NMR spectra line shape analysis.

|  | Hydroxylated UiO-66 | | Dehydroxylated UiO-66 | |
| --- | --- | --- | --- | --- |
|  | $\Delta H$, kJ mol$^{-1}$ | $\Delta S$, J mol$^{-1}$ K$^{-1}$ | $\Delta H$, kJ mol$^{-1}$ | $\Delta S$, J mol$^{-1}$ K$^{-1}$ |
| isobutane-$d_1$ | 5.3 ± 1.3 | 34 ± 10 | 5.4 ± 0.7 | 49 ± 5 |
| *n*-butane-$d_{10}$ | 1.6 ± 0.8 | -2 ± 6 | 2.9 ± 0.3 | 17.5 ± 2.3 |

To determine the rates of the motions we have measured the rates of spin relaxation as a function of the temperature. Figures 11a and 11b show the corresponding relaxation curves of *n*-butane and isobutane. Both isomers show a qualitatively similar behavior with well-resolved characteristic minima on the $T_1$ and $T_2$ relaxation curves.

Following the line shape analysis, we can conclude, that the state **I** represents the molecules confined in the smaller tetrahedral cages, while the state **II** is populated by molecules residing in the large octahedral cages. The molecules in both states are involved in a diffusion over the porous network of UiO-66 (Zr). Sequential jump exchange of the molecules among the cages represents eventually the translational jump-diffusion, which affects the line shape and the relaxation profiles as an isotropic reorientation with a correlation time, corresponding to diffusional motion [14]. The motion mechanism chosen for the fitting the relaxation data is taken without changes from our previous paper [16].

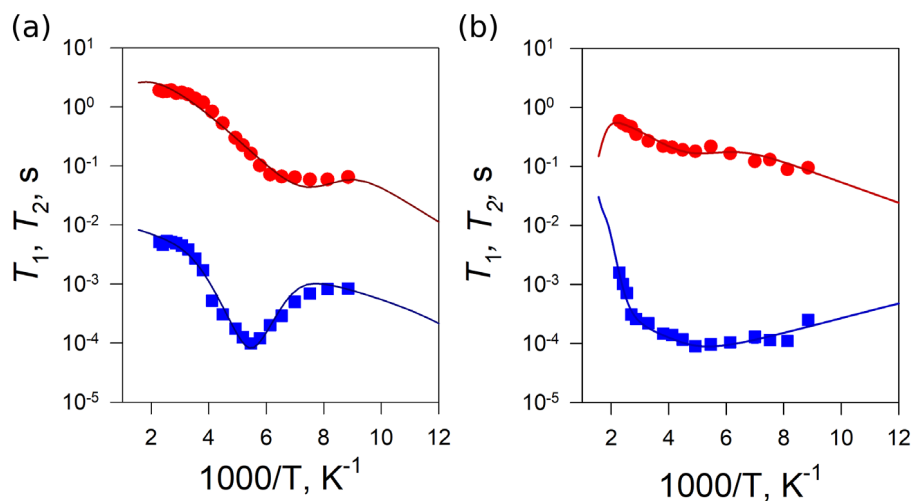

**Figure 11.** Temperature dependence of spin relaxation times for (a) *n*-butane and (b) isobutane in hydroxylated form of UiO-66 (Zr). Experimental data are presented with symbols $T_1$ (•) and $T_2$ (■), fitting is shown with solid lines. Individual relaxation times of dynamical state I (green) and II (pink) are shown with dashed lines.

Table 3 shows the kinetic parameters of butane motions in both forms of UiO-66 (Zr). Although, the geometry and types of motions are same for hydroxylated and dehydroxylated form, some kinetic parameters (activation energies and pre-exponential factors) change upon hydroxylation. The main difference between hydroxylated and dehydroxylated form of UiO-66 is manifested in the parameters of isotropic rotation ($k_{iso}$) and diffusion ($k_D$). Surprisingly, the activation barrier of isotropic motion is slightly lower in the hydroxylated form (9.5 kJ mol$^{-1}$ vs 10.5 kJ mol$^{-1}$). So, the introduction of additional interaction centers does not lead to the hindering the isotropic rotation. Apparently, the hydrogen bond energy between *n*-butane and OH-group is lower than the interaction between the guest molecules with the walls of the cage that produces the barrier for isotropic rotation.

The activation barrier $E_D$ and pre-exponential factor $k_0^D$ of the motion that we ascribe to activated diffusion actor of *n*-butane diffusion are lower in the hydroxylated form of UiO-66 (Table 3). Using Einstein equation $D = \frac{\langle l^2 \rangle}{6\tau}$, where $l \approx 1$ nm is a mean jump length between the cages and $\tau = 1/(2\pi k^D)$ we can assess the diffusivity of adsorbed species. The comparison shows, that already above 153 K the of *n*-butane diffuses faster in the dehydroxylated form. At 300 K in the hydroxylated UiO-66 the diffusion coefficient $D_n = 9.2 \times 10^{-13}$ m$^2$ s$^{-1}$ which is ~ 3 times smaller than in dehydroxylated case.

That result is in a good agreement with the data on the mobility of terephthalate linkers of UiO-66 (Zr). We have shown that the activation barrier of the flipping motion is also lower in the hydroxylated form. Therefore, the guest molecule requires less energy in order to rotate the linker and pass the window. At the same time the effective diameter of the window between adjacent cages is smaller for the hydroxylated form. That results in lower number of attempts to pass the window and diffusivity drops down ($D_{nbut}^{dehydro}/D_{nbut}^{hydro} = 6.6$ at 300 K).

**Table 3.** Kinetic parameters of all motions of butane isomers in hydroxylated and dehydroxylated form of UiO-66 (Zr).

|  | Dehydroxylated UiO-66 | | Hydroxylated UiO-66 | |
| --- | --- | --- | --- | --- |
|  | isobutane | $n$-butane | isobutane | $n$-butane |
| $E_{methyl}$, kJ mol$^{-1}$ | – | 6 |  | 6 |
| $k_0^{methyl}/(2\pi)$, s$^{-1}$ | – | 8×10$^{12}$ |  | 4×10$^{12}$ |
| $E_{cone}$, kJ mol$^{-1}$ | 6.5 | 2.5 |  | 2 |
| $k_0^{cone}/(2\pi)$, s$^{-1}$ | 1.5×10$^{10}$ | 5×10$^6$ |  | 3×10$^6$ |
| $E_{iso}$, kJ mol$^{-1}$ | 5.6 | 10.5 |  | 9.5 |
| $k_0^{iso}/(2\pi)$, s$^{-1}$ | 3.8×10$^6$ | 7×10$^{11}$ |  | 5×10$^{11}$ |
| $E_D$, kJ mol$^{-1}$ | 50 | 25 |  | 20 |
| $k_0^D/(2\pi)$, s$^{-1}$ | 1×10$^{12}$ | 5×10$^{10}$ |  | 10$^9$ |
| $E_1$, kJ mol$^{-1}$ | 6.5 | – |  | – |
| $k_{01}/(2\pi)$, s$^{-1}$ | 4×10$^{12}$ | – |  | – |
| $E_2$, kJ mol$^{-1}$ | 6.5 | – |  | – |
| $k_{02}/(2\pi)$, s$^{-1}$ | 8×10$^{12}$ | – |  | – |
| $E^{ex}$, kJ mol$^{-1}$ | 0.5 | 1.5 |  | 1.5 |
| $k_0^{ex}/(2\pi)$, s$^{-1}$ | 10$^3$ | 10$^6$ |  | 10$^6$ |

Notably, the isobutane diffusion remains almost the same with diffusion coefficient $D_{iso}$ = 4.9×10$^{-15}$ m$^2$ s$^{-1}$ at 300 K. Thus, in the hydroxylated framework, the difference in diffusion coefficients is only 185, more than 5 time worse compared to the hydroxylated case. We can conclude, that if we want to preserve the separation selectivity, we need to (i) use the dehydroxylated materials and (ii) pre-condition the gas mixture to remove water traces, as UiO-66 is known to turn back to the hydroxylated form in humid conditions.

### 4. Conclusion

The molecular mobility of terephthalate linkers in UiO-66 and the dynamics of adsorbed butane isomers have been characterized by $^2$H NMR method. Comparison of motional parameters in hydroxylated and dehydroxylated form of UiO-66 reveals that the activation barrier of $n$-butane diffusion and its diffusivity are lower in the hydroxylated form of UiO-66 whereas the diffusion of isobutane is not affected at all. It leads to the higher separation selectivity of butane isomers in dehydroxylated form of UiO-66. This result can be rationalized by the influence of the hydroxylation on the linker mobility. Lower activation barrier of the linker flipping motion leads to the lower activation barrier of diffusion. Smaller effective diameter of the window between cages and possibility of forming hydrogen bond between the guest molecule and OH-group lead to the lower number of attempts to cross the window. Therefore, the diffusivity drops down despite the advantage of lower activation barrier of diffusion.

**AUTHOR INFORMATION**

**Corresponding Authors**

*E-mail: kdi@ catalysis.ru (D.I.K.).

*E-mail: stepanov@catalysis.ru (A.G.S.); Tel : +7 9529059559 ; Fax: +7 383 330 8056

**Author Contributions**

The manuscript was written through contributions of all authors.

**Notes**

The authors declare no competing financial interests.

**ACKNOWLEDGMENT**

This work was supported by Russian Foundation for Basic Research (RFBR) (Grant no. 19-33-90026).